
\input amstex
\documentstyle{amsppt}
\hoffset=0.75in
\voffset=0.5in
\NoBlackBoxes
\leftheadtext{R. I. MCLACHLAN}
\rightheadtext{EXPLICIT LIE-POISSON INTEGRATION AND THE EULER EQUATIONS}
\topmatter

\title Explicit Lie-Poisson integration and the Euler equations
\endtitle
\date April 27, 1993\enddate
\author Robert I. McLachlan\endauthor
\address Program in Applied Mathematics, University of Colorado at Boulder,
Boulder, CO 80309-0526
\endaddress
\email rxm\@boulder.colorado.edu\endemail

\font\sym=msam10

\font\bb=msym10
\def\RR {\hbox{\bb R}}
\font\mib=cmmib10
\def\bld#1{\hbox{\textfont1=\mib$#1$}}

\def\z{\bold z}

\def\n{{\bold n}}
\def\m{{\bold m}}
\def\j{{\bold j}}
\def\k{{\bold k}}

\def\dt{\Delta t}
\font\german=eufm10
\def\gg#1{\hbox{\german #1}}
\def\eps{\varepsilon}
\def\to{\!\!\rightarrow\!\!}


\abstract
We give a wide class of Lie-Poisson systems for which explicit, Lie-Poisson
integrators, preserving all Casimirs, can be constructed. The integrators are
extremely simple. Examples are the rigid body, a moment truncation,
and a new, fast algorithm for the sine-bracket truncation of the 2D Euler
equations.
\endabstract

\endtopmatter

\document

\head \S1. Introduction
\endhead                             

Hamiltonian systems are fundamental, and symplectic integrators (SI's) have
been increasingly used to do useful extremely-long-time numerical
integrations of them. Wisdom \cite{17} has used fast SI's to integrate
the solar system far more efficiently than with standard methods; there
are numerous examples illustrating the superior preservation of phase-space
structures and qualitative dynamics by SI's \cite{9,10,11,14}. Many Hamiltonian
systems are not in canonical form but are most naturally written as
{\it Lie-Poisson} systems, which generally arise as reductions from
canonical formulations in more variables. Examples are rigid bodies,
fluid-particle-field systems (e.g. magnetohydrodynamics or the
Vlasov-Poisson equations \cite{7,8}) and general relativity.

Integrating
Lie-Poisson PDE's first requires a truncation of the underlying Lie
algebra, the most promising approaches at this time being the sine bracket
(\cite{5}, associated  with $\gg{su}(N)$ and applying to systems
whose configuration space is the set of symplectic maps of a $2n$-%
dimensional torus, such as 2D incompressible
fluids) and, for localized distributions, the moment truncation of
Scovel and Weinstein \cite{15}. Then, to integrate in time, there
are general methods which provide a symplectic-leaf-preserving Poisson map
\cite{4,6}. They are not only implicit but require evaluating functions
like ``$e^{ad_\xi}$'' via Taylor series; hence they can be very slow.
Although
they can be of any order \cite{2}, the time-step must be kept small so that
the implicit equations can be solved quickly by iteration---so the implicitness
is no advantage.

If Lie-Poisson integrators are to be as practical as standard symplectic
integrators they should be simple and fast.
To this end we describe the widest general class of such systems for which
explicit methods are available.
(Examples of such methods were first found by Ruth \cite{13} for canonical
systems, and by Channell and Scovel \cite{2} for Lie-Poisson systems.)
Our class includes the sine-bracket truncation
of the 2D Euler equations---the {\it sine-Euler} equations---for which
the new method is not only explicit but $\Cal O(N/\log N)$ times
faster than the standard implicit method.

\head \S2. Poisson systems \& integrators
\endhead
A Lie-Poisson system has a phase space $M\cong\RR^n\ni x$, a Lie-Poisson
bracket $\{F,G\}={\partial F\over\partial x_i}J_{ij}{\partial G\over
\partial x_j}$ where $J_{ij}=c_{ij}^k x_k$ (with $c_{ij}^k$ the
structure constants of a Lie algebra)
and a Hamiltonian $H\!\!\!:M\to\RR$. The dynamics
$\dot x=\{x,H\}=J\nabla H$
preserve the Poisson bracket;
a Poisson integrator is one
whose time-step map $x\to x'(x)$ also preserves the Poisson bracket.
Symplectic splitting methods apply when the Hamiltonian is a sum of terms
each of which can be explicitly integrated---for example, in a
canonical Hamiltonian system, $H=T(p)+V(q)$, which leads to standard
symplectic integrators \cite{13,14}. For Lie-Poisson systems, our methods
depend
on the following observation. We first form the set of all abelian
subalgebras of $M$.

\proclaim{Observation}
Let
$$\Sigma = \{\sigma_k\subset\{1,\dots n\}: J_{ij}=0\quad\forall\
i,j\in\sigma_k\}$$
Then the
Lie-Poisson system with Hamiltonian $H(\sigma_k)$ (i.e. $H$ depends only
on $x_i$ with $i\in\sigma_k$) is linear with constant
coefficients.
\endproclaim

So if these linear systems can be solved exactly, an explicit first-order
Poisson integrator for a Hamiltonian with $p$ terms
$H=\sum_{k=1}^p H_k(\sigma_k)$ is
$$\varphi= \exp(\dt X_1)\dots \exp(\dt X_p)\eqno(1)$$
where $\dt$ is the time step and $X_k=J\nabla H_k$; that is, just integrate
each piece of the Hamiltonian in turn.
A second order symmetric method (``leapfrog'') is
$\varphi({1\over2}\dt)\varphi^{-1}(-{1\over2}\dt) =$
$$\exp({\dt\over2} X_1)\ldots \exp({\dt\over2} X_{p-1})\exp(\dt X_p)
\exp({\dt\over2} X_{p-1})\ldots \exp({\dt\over2} X_1).$$
Methods of any order can be constructed by composing several such steps
\cite{2,9,16,18}. Clearly an arbitrary linear term could also be included in
any
of the $H_k$.

Now $\Sigma$ certainly includes the singleton subsets so we can immediately
integrate Hamiltonians of the form $H=\sum_{k=1}^n H_k(x_k)$. We choose
not to break with tradition and include the rigid body as an example.

\example{Example: The rigid body}
Here $\m\in\RR^3$ is the angular momentum, $H={1\over2}({m_1^2\over I_1}
+{m_2^2\over I_2}+{m_3^2\over I_3})$, and the Lie algebra is $\gg{so}(3)$
so
$$J=\pmatrix 0&-m_3&m_2\cr m_3&0&-m_1\cr-m_2&m_1&0\endpmatrix.$$
This $J$ has a Casimir $C=|\m|^2$ (the total angular momentum) so
$\widehat H=H-C/2I_1$ has the same dynamics; this leaves only two terms
in $\widehat H$.
With $\omega_k=m_k({1\over I_k}-{1\over I_1})$ and $R_k(\theta)$
rotation by $\theta$ around the axis $m_k$,
the map corresponding to (1) is
$$ \m'=R_3(\dt\omega_3)R_2(\dt\omega_2)\m $$
---a ``standard map'' of the rigid body. The symplectic leaves $|\m|^2=$ const
are clearly preserved. (We believe that this simple method (extended to
higher order) would improve the rigid body simulations
of Austin {\it et al.} \cite{1} who used the (implicit, non-Poisson) midpoint
rule.)
The same technique applied to the planar pendulum
in the form
$$H={1\over2}x_3^2+x_2,\qquad J=\pmatrix 0&0&-x_2\cr0&0&x_1\cr
x_2&-x_1&0\endpmatrix$$
($x_1$, $x_2$ coordinates in the plane, $x_3=$
angular velocity; Casimir $x_1^2+x_2^2=$ length of pendulum)
does in fact give the standard map (see, e.g. \cite{4}).
\endexample

Even if $H$ is of the form $\sum H_k(\sigma_k)$,
these methods are only practical if
the resulting linear systems can be integrated exactly---diagonalizing
them numerically would destroy their efficiency. This has not been
a problem in the cases we have done to date: often the linear system
is already upper triangular (after a permutation) and can be
solved by back-substitution.

\def\la{\langle}
\def\ra{\rangle}
\example{Example: A moment algebra}
Here $M$ is the algebra of 2nd and 4th order moments of a distribution
$f(q,p)$. Coordinates are $\la q^\alpha p^\beta\ra\equiv\int q^\alpha
p^\beta f(q,p)\,dqdp$ which we collect in a vector $x$.
It could arise as a truncation of the evolution of $f(q,p)$ by
Hamiltonian vector fields on the plane, such as the Liouville equation
$\dot f+\{f,h\}_{qp}=0$, the 2D Euler equations, or the 1D
Vlasov-Poisson equations \cite{3,16}. We have
$$x=\left(\!\!\matrix\la q^2\ra\cr\la qp\ra\cr\la p^2\ra\cr\la q^4\ra\cr\la
q^3p\ra\cr
\la q^2p^2\ra\cr\la qp^3\ra\cr\la p^4\ra\endmatrix\!\!\right),\quad
J=\pmatrix
0&2x_1&4x_2&0&2x_4&4x_5&6x_6&8x_7\cr
-2x_1&0&2x_3&-4x_4&-2x_5&0&2x_7&4x_8\cr
-4x_2&-2x_3&0&-8x_5&-6x_6&-4x_7&-2x_8&0\cr
0&4x_4&8x_5&0&0&0&0&0\cr
-2x_4&2x_5&6x_6&0&0&0&0&0\cr
-4x_5&0&4x_7&0&0&0&0&0\cr
-6x_6&-2x_7&2x_8&0&0&0&0&0\cr
-8x_7&-4x_8&0&0&0&0&0&0\cr\endpmatrix
$$
As pointed out in \cite{3}, if $H$ were separable in $p$ and $q$ as in
the Vlasov-Poisson equations, i.e.
$H=T(x_3,x_8)+V(x_1,x_4)$, then an explicit Poisson integrator is
possible, just as in the canonical case. But from the above observation
is it clear that
$$H=H_1(x_1,x_4)+H_2(x_2,x_6)+H_3(x_3,x_8)+H_4(x_4,x_5,x_6,x_7,x_8)$$
also admits an explicit Poisson integrator.
\smallskip
\noindent{\it Extra nonlinear terms--- } In the canonical case one often
has a nonlinear term in $H$, not a function of $q$ alone, which can
nevertheless be integrated explicitly (and conveniently, i.e. without
using nonelementary functions). One such is $H(q_1,q_2^2+p_2^2)$, which
arises in the nonlinear Schr\"odinger equation and in the Zakharov
equations \cite{9} and which gives rise to the well-known splitting method
for the nonlinear Schr\"odinger equation. This phenomenon is less common
in Lie-Poisson systems because of the more complicated evolution of those
$x_j$ not appearing in $H$. But it can happen, e.g. let $H=H(x_1x_3)$ in the
above algebra:
$$\eqalign{
\dot x_1&=\lambda x_1\cr
\dot x_2&=0\cr
\dot x_3&=-\lambda x_3\cr
}\quad\Rightarrow \quad\eqalign{
x_1(t)&=e^{\lambda t}x_1(0)\cr
x_2(t)&=x_2(0)\cr
x_3(t)&=e^{-\lambda t}x_3(0)\cr}
$$
$$\pmatrix
\dot x_4\cr \dot x_5\cr\dot x_6\cr\dot x_7\cr \dot x_8\endpmatrix
=\pmatrix
0&8e^{\lambda t}b&0&0&0\cr
-2e^{-\lambda t}a&0&6e^{\lambda t}b&0&0\cr
0&-4e^{-\lambda t}a&0&4e^{\lambda t}b&0\cr
0&0&-6e^{-\lambda t}a&0&2e^{\lambda t}b\cr
0&0&0&-8e^{-\lambda t}a&0&\cr\endpmatrix
\pmatrix x_4\cr x_5\cr x_6\cr x_7\cr x_8\endpmatrix
$$
where $a=x_3(0)$, $b=x_1(0)$, and $\lambda=4x_2(0)H'(ab)$ is constant.
This time-dependent linear system can be solved: let $L(t)$ diagonalize
its matrix; then because $L^{-1}\dot L$ is time-independent and
tridiagonal, the
system is transformed by $L$ to a time-independent, tridiagonal one, which can
be solved by standard methods.
\endexample

It is not always advantageous to split $H$ into the fewest possible number of
pieces. Splitting even further can lead to less coupled systems to solve
and does not affect the error.

The local truncation error of such methods can be easily derived using
the Campbell-Baker-Hausdorff formula (see \cite{9}). For a method of order
$l$, it is a sum of all iterated Poisson brackets of order $l+1$ of the
$H_k$'s, with coefficients depending on the order in which the $H_k$'s
are taken in (1). It may look as if the error ``increases with the number of
pieces in $H$''; but
this would mean that the number of ODE's is also increasing, which will
affect the error of any method. In fact, numerical simulations show
that these methods have roughly the same truncation errors as
the implicit Poisson methods.
For example, for the rigid body, the
above method was $1.6\times$ less accurate but $100\times$ faster than
the implicit method.

\head \S3. The sine-Euler equations
\endhead

Here we consider the motion of an inviscid incompressible fluid governed
by the 2D Euler equations. The field variable is the vorticity $\omega(x,y)$,
which is $2\pi$-periodic in $x$ and $y$. The phase space of vorticities is
the dual of the Lie algebra of Hamiltonian vector fields in the plane
\cite{8}. We have
$$ J=\omega_y\partial_x-\omega_x\partial_y\eqno(2)$$
$$\eqalign{
H&=-{1\over2}\int\psi\omega\,dxdy \quad\hbox{where\ }\nabla^2\psi=-\omega\cr
\dot w&=J(\omega){\delta H\over\delta\omega} = J(\omega)\psi =
-\omega_x\psi_y+\omega_y\psi_x}\eqno(3)$$
This algebra has an infinite number of Casimirs, which we may take as
$$C_n=\int\omega^n\,dx\,dy$$
In Fourier space eqs. (2,3) become
$$\eqalign{
J_{\m\n}&=\m\times\n\,\omega_{\m+\n}\cr
H&={1\over2}\sum_{\n\ne{\bold 0}}{\omega_\n\omega_{-\n}\over|\n|^2}\cr
\dot\omega_\m&=\sum_{\n\ne{\bold 0}}
{\m\times\n\over|\n|^2}\omega_{\m+\n}\omega_{-\n}}$$
where $\m\times\n=m_1 n_2 - m_2 n_1$ and for real $\omega$,
$\omega_{-\n}=\omega_\n^*$.
There is a finite-dimensional truncation of (2) \cite{5}, the {\it sine
bracket}
$$ J_{\m\n}={1\over\eps}\sin(\eps\m\times\n)\omega_{\m+\n\mod N}\eqno(4)$$
where $\eps=2\pi/N$ and all indices are henceforth reduced modulo $N$ to the
periodic lattice $-M\le m\le M$ where $N=2M+1$, and we shall take $N$ prime
for convenience (the extensions to nonprime and even $N$ are straightforward).
This $J$ has $N-1$ Casimirs which approximate $C_n$ for $2\le n\le N$.
Eqs. (4) specify the structure constants of $\gg{su}(N)$ in an appropriate
basis, see \cite{5,19}.

The most natural truncated $H$ is
$$ H={1\over2}\sum_{\scriptstyle n_1,n_2=-M\atop\scriptstyle \n\ne{\bold 0}}^M
{\omega_\n\omega_{-\n}\over |\n|^2}$$
giving the sine-Euler equations, first proposed by Zeitlin \cite{19}:
$$ \dot\omega_\m=\sum_{\scriptstyle n_1,n_2=-M\atop\scriptstyle \n\ne{\bold
0}}^M
{1\over\eps}{\sin(\eps\m\times\n)\over|\n|^2}\omega_{\m+\n}\omega_{-\n}
\eqno(5)$$
where as in (4) all indices are taken modulo $N$.
As a numerical approximation of (3), these equations are only $\Cal O(\eps^2)$
accurate. However, once the vorticity has rolled up into small scales, standard
(finite difference or spectral) approximations are not very accurate either,
and do not possess the Poisson structure or conserved quantities of (5).
Perhaps (5) is best regarded as an extremely interesting {\it model} of
the 2D Euler equations until its properties are better understood.

The approach outlined above now gives an $\Cal O(N^3\log N)$, explicit
Poisson integrator of (5), preserving all $N-1$ Casimirs to within
round-off error---which is faster than the $\Cal O(N^4)$ needed just
to evaluate the right hand side of (5). We take
$$\sigma_\k=\{n\k: 0\le n<N\}\in\Sigma.$$
To split the Hamiltonian we need a set of modes that generates the entire
lattice, e.g.
$$ K=\{(0,1)\}\cup\{(1,m): 0\le m<N\}$$
(this is why it is convenient to take $N$ prime---otherwise multiples of $K$
don't cover the entire lattice) and then
$$ H=\sum_{\k\in K}H_\k(\sigma_\k),\qquad
H_\k={1\over2}\sum_{n=0}^{N-1}{\omega_{n\k}\omega_{-n\k}\over|n\k|^2}.$$

We now solve the linear ODE's $\dot\omega=J\nabla H_\k$. Of course
$\dot\omega_\m=0$ for $\m\in\sigma_\k$. From (5), the other modes decouple
into $2M$ sets of $N$ equations, of which we only need to solve $M$ sets; the
others are their complex conjugates (see Figure 1).
The variables in each set are a
translation of $\sigma_\k$, say by $\j$: let $z_m=\omega_{\j+m\k}$, then
$$\dot z_m=\sum_{n=-M}^M a_n z_{m-n}$$
where
$$ a_n= -{\sin(\eps n \j\times\k)\over\eps|n\k|^2}\omega_{n\k}$$
These ODE's are circulant and hence diagonalized by the discrete Fourier
transform: let $\widetilde\z=F\z$ where $F$ is the DFT; then
$$ \dot{\widetilde \z}=\Lambda \widetilde\z\hbox{\quad where\quad }
\Lambda=\hbox{diag}(F{\bold a}) $$
so the equations can now be integrated explicitly.

\def\itemitemitem{\par \indent\indent \hangindent 3\parindent \textindent}
\example{Summary of algorithm}
\item{$\bullet$} for $\k\in K$ do
\itemitem{$\bullet$} for $\j=1^{\hbox{st}},\dots M^{\hbox{th}}$
translation of $\k$ do [may be done in parallel]
\itemitemitem{$\bullet$} with $z_m=\omega_{\j+m\k}$, set
$\z'=F^{-1}e^{\dt\Lambda}F\z$
\itemitemitem{$\bullet$} copy $(\z')^*$ into $\omega_{-(\j+m\k)}$
\itemitem{$\bullet$}end do
\item{$\bullet$}end do
\endexample

The whole procedure requires $3M(N+1)={3\over2}(N^2-1)$ DFT's of length $N$.
As there are FFT's available for sequences of prime length $N$ \cite{12}, the
whole algorithm is $\Cal O(N^3\log N)$. Figure 2 shows the relative energy
error $(H(t)-H(0))/H(0)$ for $10^5$ time steps with $\dt=0.05$, $N=7$,
$|\bld{\omega}|=1$
and $H(0)=0.75$. As is expected from an integrator which is a symplectic map
on the symplectic leaves of the phase space, the energy error does not
grow with time. The errors in the Casimirs, e.g.
$C_2=\sum\omega_\n\omega_{-\n}$
are due only to roundoff error, and grow by about $5\times10^{-15}$ per
time step.

This explicit method exists because the only coupled terms in the
Hamiltonian belong to the sets $\sigma_\k$; it is fast because of the
special form of the sine bracket in this basis.
Preliminary simulations indicate that the vorticity does not roll up into
the high modes and that the evolution can be followed for arbitrarily long
times. It will be interesting to see what the implications of this model
are for the ergodicity and statistical steady state of the 2D Euler
equations.

\bigskip
\noindent{\it Acknowledgements.} I would like to thank Clint Scovel for
a careful reading of the manuscript, and for encouraging to me to struggle,
to seek, to find and not to yield to Poissonology.

\pageinsert
\hbox{ }\vskip 3in
{\smc Figure 1.} Mode splitting in the sine-Euler equations. Here $N=7$ and
$M=3$.
$\bigcirc$ shows modes in one term $H_\k$ in the Hamiltonian ($\k=(1,1)$).
{\sym\char'004 } shows modes which are coupled together in the linear system
$\dot\omega=J\nabla H_\k$
(here $\j=(2,0)$); {\sym\char'003 } shows modes whose values are the complex
conjugate of the {\sym\char'004 } modes.
\vskip 3in
\centerline{{\smc Figure 2.} Relative energy error to $t=5000$, $\dt=0.05$,
$N=7$.}
\endinsert

\enddocument